\documentstyle[preprint,aps,epsfig,eqsecnum,tighten]{revtex} 
\catcode`@=11
\def\references{%
\ifpreprintsty
\bigskip\bigskip
\hbox to\hsize{\hss\large \refname\hss}%
\else
\vskip 24pt
\hrule width\hsize\relax
\vskip 1.6cm
\fi
\list{\@biblabel{\arabic{enumiv}}}%
{\labelwidth\WidestRefLabelThusFar  \labelsep4pt %
\leftmargin\labelwidth %
\advance\leftmargin\labelsep %
\ifdim\baselinestretch pt>1 pt %
\parsep  4pt\relax %
\else %
\parsep  0pt\relax %
\fi
\itemsep\parsep %
\usecounter{enumiv}%
\let\p@enumiv\@empty
\def\theenumiv{\arabic{enumiv}}%
}%
\let\newblock\relax %
\sloppy\clubpenalty4000\widowpenalty4000
\sfcode`\.=1000\relax
\ifpreprintsty\else\small\fi
}
\catcode`@=12
\draft
\newcommand{\nc}{\newcommand}
\nc{\beq}{\begin{equation}}   \nc{\eeq}{\end{equation}}
\nc{\bea}{\begin{eqnarray}}   \nc{\eea}{\end{eqnarray}}
\nc{\baa}{\begin{array}}      \nc{\eaa}{\end{array}}
\nc{\bit}{\begin{itemize}}    \nc{\eit}{\end{itemize}} 
\nc{\ben}{\begin{enumerate}}  \nc{\een}{\end{enumerate}}
\nc{\bce}{\begin{center}}     \nc{\ece}{\end{center}}
\def\beqa{\begin{eqnarray}}
\def\eeqa{\end{eqnarray}}
\def\be{\beq}
\def\ee{\eeq}
\def\to{\rightarrow}

\def\nmu{\nu_\mu}

\def\lsim{\mathrel{\raise.3ex\hbox{$<$\kern-.75em\lower1ex\hbox{$\sim$}}}}
\def\gsim{\mathrel{\raise.3ex\hbox{$>$\kern-.75em\lower1ex\hbox{$\sim$}}}}

\begin{document}
\draft
\tightenlines
\topmargin = -1.0cm
\overfullrule 0pt
\preprint{\vbox{\hbox{BU-01-26}
\hbox{CSUDH-EPRG-01-05}
\hbox{IFUSP-DFN/01-061}
\hbox{MADPH-01-1242}
\hbox{hep-ph/0110393}
}}
\title{
Neutrino oscillation parameters from \\
MINOS, ICARUS and OPERA combined}
\author{ 
V.\ Barger$^1$,
A.\ M.\ Gago$^{2,3,4}$,
D.\ Marfatia$^{1,5}$,
W.\ J.\ C.\ Teves$^3$,\\ 
B.\ P.\ Wood$^1$ 
and R.\ Zukanovich Funchal$^3$}
\address{\sl 
$^1$ Department of Physics, University of Wisconsin-Madison, WI 53706, USA\\
$^2$ Department of Physics, California State University, Dominguez Hills, Carson 
CA 90747, USA\\
$^3$ Instituto de F\'{\i}sica, Universidade de S\~ao Paulo 
C.\ P.\ 66.318, 05315-970 S\~ao Paulo, Brazil\\
$^4$  Secci\'{o}n F\'{\i}sica, Departamento de Ciencias, Pontificia
Universidad Cat\'olica del Per\'u, \\Apartado 1761, Lima, Per\'u\\
$^5$ Department of Physics, Boston University, Boston, MA 02215, USA\\
}
\maketitle
\vspace{-0.4cm}
\hfuzz=25pt
\begin{abstract}
We perform a detailed analysis of the capabilities of the MINOS, 
ICARUS and OPERA experiments to measure neutrino oscillation parameters 
at the atmospheric scale with their data taken separately and in combination. 
MINOS will determine $\Delta m^2_{32}$ and $\sin^2 2\theta_{23}$
to within 10\% at the 99\% C.L. with 10 kton-years of data.
While no one experiment will determine $\sin^2 2\theta_{13}$ with much precision, 
if its value lies in the combined sensitivity region of the three
experiments, it will be possible to place a 
{\it lower bound} of ${\cal{O}}(0.01)$ at the \mbox{95\% C.L.} 
on this parameter by combining the data from the three experiments.
The same bound can be placed with a combination of MINOS and ICARUS data alone.

\end{abstract}

\newpage

\section{Introduction}
\label{sec:intro}

The hypothesis that neutrino oscillations are responsible for the 
atmospheric~\cite{kam,imb,macro,soudan,SKatm} and
solar~\cite{homestake,kamsolar,gallex,gno,sage,sk,sno}
 neutrino anomalies is becoming increasingly 
favored as more data is accumulated and new experiments start running.
In particular, the disappearance of atmospheric $\nmu$ is 
currently being tested by K2K, the first long baseline experiment, with
confirming indications~\cite{K2K}. The best-fit oscillation parameters relevant to
the atmospheric and solar scales from the analysis of the data from the above experiments are
$(\Delta m^2_{32}, \sin^2 2\theta_{23})=(2.9 \times 10^{-3}\, {\rm{eV}}^2, 0.98)$~\cite{fogli1}
and $(\Delta m^2_{21}, \sin^2 2\theta_{12})=(4.9\times 10^{-5}\, {\rm{eV}}^2, 0.79)$~\cite{fogli2},
respectively. Similar results were obtained in 
other global analyses of solar data; see Ref.~\cite{othersol}. 
The remaining mixing angle, $\theta_{13}$, is constrained by the CHOOZ
experiment to be small, $\sin^2 2\theta_{13} \lsim 0.1$, for values of $\Delta m^2_{32}$ 
relevant to atmospheric neutrino oscillations at the 95\% C.L.~\cite{chooz}.

If oscillations in the three neutrino framework are assumed to be the correct solution 
to the solar and atmospheric neutrino problems (although the participation
of a sterile neutrino in the oscillation dynamics, motivated by the LSND~\cite{lsnd} data, 
is convincingly consistent with all existing data~\cite{sterile}), the mission of future
experiments will be to precisely measure the relevant mixing parameters. 
The question of how well this task can be accomplished by the upcoming 
long baseline neutrino facilities is a rather important one since it 
will guide studies and set the goals for future neutrino experiments.

The KamLAND reactor experiment~\cite{kamland} will take important steps to 
solve the solar neutrino problem 
by determining if the large mixing angle MSW solution is correct, and by pinning down 
$\Delta m^2_{21}$ and $\sin^2 2\theta_{12}$ to accuracies of 
${\cal{O}}(10)$\% if it is~\cite{bdw}. 

 Low statistics in the K2K experiment will limit its 
power to measure the oscillation parameters relevant to the atmospheric neutrino problem. 
Three other long baseline accelerator neutrino experiments will  
address the atmospheric neutrino deficit with much higher statistics: 
the Main Injector Neutrino Oscillation Search (MINOS) 
experiment~\cite{minos,minos1,minos2} from FNAL to the Soudan Mine, 
 the Imaging Cosmic And Rare Underground Signals
(ICARUS) experiment~\cite{icarus,icarus2} and the 
Oscillation Project with Emulsion-tRacking Apparatus 
(OPERA)~\cite{opera,opera2} from CERN 
to Gran Sasso. All are expected to start taking data around 2005, which
will be long before the advent of superbeams~\cite{sup} or neutrino factories~\cite{geer};
for more references see Ref.~\cite{E1}.
These future long baseline experiments and KamLAND will therefore provide 
independent and accurate information about neutrino oscillation parameters 
at the solar
and atmospheric scales with the important advantage that their results
will not rely on assumptions about the solar model or the atmospheric 
neutrino fluxes. 

In this work we explore
the precision with which $\Delta m^2_{32}$, $\sin^2 2\theta_{23}$ and $\sin^2 2\theta_{13}$ 
can be measured by MINOS, ICARUS and OPERA in the framework of three neutrino oscillations.  
We include all oscillation channels accessible to 
these experiments and take 
into account the available experimental information on
backgrounds and efficiencies. 
All three experiments will operate with a high $\nu_\mu$ flux and measure $\nu_\mu$ disappearance. 
Because of the energy dependence of the oscillation probability, the most sensitive measure
of the oscillation parameters is the energy spectra of the neutrinos  
reaching the detector. 
Observations of an energy dependent distortion of the spectrum can 
uniquely  determine the oscillation parameters. Charged current events 
alone can provide information about the original $\nu_\mu$ energy spectrum.

Although these experiments can also observe $\nu_e$, and to a lesser extent $\nu_\tau$ 
appearance, the expected statistics in these channels are unlikely to be high enough to allow
the energy spectra of these channels to be studied extensively. One can nevertheless extract 
important information from the total expected rates in these channels.

The paper is organized as follows. In Sec.~\ref{sec:form}
we very briefly review the three generation oscillation 
formalism. 
In Secs.~\ref{sec:minos} and~\ref{sec:opera} we describe the MINOS, ICARUS,
and OPERA experiments defining the observables relevant to our analyses and 
how they are calculated. In Sec.~\ref{sec:ana} we outline the procedure 
 used to
analyse the data simulated for MINOS, ICARUS and OPERA. We also perform 
 a global analysis of the data from the three
experiments. In Sec.~\ref{sec:res} 
we discuss our results on the sensitivity to and precision with which 
 $\Delta m^2_{32}$, $\sin^2 2\theta_{23}$ and $\sin^2 2\theta_{13}$ are expected to be 
determined by each experiment, and from the combined data of all three experiments. 
 We present our conclusions in Sec. \ref{sec:conc}.

\section{3 $\nu$ Oscillations}
\label{sec:form}

In the three generation framework there are three mixing angles and one $CP$ phase
that determine the amplitude for transition from flavor state to another. For the 
experiments under study, the $CP$ phase has a negligible effect on the oscillation
amplitudes and we set it to zero in what follows.
The transformation between  the neutrino mass eigenstates $\nu_1$, 
$\nu_2$ and $\nu_3$ and neutrino interaction eigenstates $\nu_e$, $\nu_\mu$ 
and $\nu_\tau$ is given by
\be
\left( 
\begin{array}{c}
\nu_e \\
\nu_\mu \\
\nu_\tau
\end{array}
\right)
= U 
\left( 
\begin{array}{c} 
\nu_1 \\
\nu_2 \\
\nu_3
\end{array}
\right)
= \left[ 
\begin{array}{ccc}
c_{12}c_{13} & s_{12}c_{13} & s_{13} \\
-s_{12}c_{23}-c_{12}s_{23}s_{13} &
c_{12}c_{23}-s_{12}s_{23}s_{13} & s_{23}c_{13} \\
s_{12}s_{23}-c_{12}c_{23}s_{13} & 
-c_{12}s_{23}-s_{12}c_{23}s_{13} & c_{23}c_{13}
\end{array}
\right]
\left( 
\begin{array}{c} 
\nu_1 \\
\nu_2 \\
\nu_3
\end{array}
\right),
\label{rel}
\ee
using the standard parameterization for the 
 mixing matrix $U$~\cite{pdg}. Here  
$c_{ij}$ and $s_{ij}$ denote the cosine and the sine of the mixing angle $\theta_{ij}$.  

The propagation of neutrinos through matter~\cite{matter} 
is described by the evolution equation
\be  
i{\displaystyle\frac{d}{dr}}  \left( \begin{array}{c} \nu_e \\ \nu_\mu \\ \nu_\tau \end{array} \right) 
=  \frac{1}{2E_\nu} \left[ U \left(\begin{array}{ccc} \Delta m^2_{12} & 0 & 0 \\
0 & 0 & 0\\
0 & 0 & \Delta m^2_{32}
\end{array} \right) U^\dagger +
\left( \begin{array}{ccc} A_e(r)  & 0 & 0 \\
0 & 0 &  0\\ 
0 & 0 & 0
\end{array} \right) \right]
\left( \begin{array}{c} \nu_e \\ \nu_\mu \\ \nu_\tau  \end{array} \right),
\label{evol} 
\ee
where $\Delta m^2_{ij}=m^2_i-m^2_j$ and $E_\nu$ is the neutrino energy. 
Here, 
$A_e(r)= 2 \sqrt{2} \,G_F n_e(r) E_\nu = 1.52 \times 10^{-4} \text{ eV}^2 \, 
Y_e \,\rho \, (\text{g/cm}^3) \, E_\nu \,(\text{GeV}) $  is the 
amplitude for $\nu_e-e$ forward scattering in matter with $Y_e$ denoting the 
electron fraction and $\rho$ the matter density.
For the experiments considered in this work, the neutrino path only traverses the
Earth's crust which has an almost constant density of $\rho \sim$ 3 g/cm$^3$ with $Y_e \sim 0.5$, 
giving $A_e \sim 0.23 \times 10^{-3} \text{ eV}^2 \, E_\nu \,(\text{GeV})$.
Although analytical expressions for the oscillation probabilities in matter of constant 
density exist~\cite{3gexpr}, we will solve Eq.~(\ref{evol}) numerically taking into
account the dependence of density on depth using the density profile from the 
Preliminary Reference Earth Model~\cite{prem}.

Since the baseline of the experiments under consideration is 730 km, and  
 $\vert \Delta m^2_{21} \vert \ll \vert \Delta m^2_{32} \vert$,
the contribution to 
the neutrino dynamics from the solar scale is small and  
the oscillation probabilities can be described in terms of just three 
parameters: $\Delta m^2_{32}$, $\sin^2 2\theta_{23}$, and
$\sin^2 2\theta_{13}$. For a constant matter density one can 
approximate the probability that a muon
neutrino is converted to an electron neutrino as~\cite{matter,3gexpr}
\be
P_{\nu_\mu \to \nu_e} \approx \sin^2 2 \theta_{13}^m \sin^2 \theta_{23} \sin^2 \left(\displaystyle \frac{\Delta m_{32}^2 L}{4E} S \right)\,,
\label{analytic1}
\ee
where 
\be
S = \displaystyle \sqrt{\left( \frac{A_e}{\Delta m^2_{32}}-\cos 2 \theta_{13} \right)^2 + \sin^2 2\theta_{13}}\,,
\label{M}
\ee
and
\be
\sin^2 2\theta_{13}^m = \displaystyle \frac{\sin^2 2\theta_{13}}{\left( \frac{A_e}{\Delta m^2_{32}}-\cos 2 \theta_{13} \right)^2 + \sin^2 2\theta_{13}}\,.
\label{sinmatter}
\ee
For $L \sim 730$ km, 
matter effects make a negligible contribution to the probability for conversion into a 
tau neutrino 
and the vacuum expression serves as a good approximation,
\be
P_{\nu_\mu \to \nu_\tau} \approx \sin^2 2 \theta_{23} \cos^4 \theta_{13} 
\sin^2 (\displaystyle \frac{\Delta m_{32}^2 L}{4E} ).
\label{analytic2}
\ee
From Eq.~(\ref{analytic1}) one can see that the 
conversion probability for $\nu_\mu \to \nu_e$ 
is approximately proportional to $\sin^2 2 \theta_{13}$ and is therefore small  
unless there is a resonant enhancement at
$\cos 2\theta_{13} = A_e/\Delta m^2_{32}$. 
The  $\nu_\mu \to \nu_\tau$ conversion probability is, on the other 
hand, proportional to $\sin^2 2 \theta_{23}$, and is therefore
large. The survival probability, 
$P_{\nu_\mu \to \nu_\mu} = 1 - P_{\nu_\mu \to \nu_\tau} - P_{\nu_\mu \to \nu_e}$ 
depends almost entirely on $\sin^2 2 \theta_{23}$ and $\Delta m^2_{32}$.
Thus the parameters $\sin^2 2 \theta_{23}$ and 
$\Delta m^2_{32}$ can be studied via $\nu_\mu$ disappearance or $\nu_\tau$ appearance, while
$\nu_e$ appearance is necessary to probe $\sin^2 2 \theta_{13}$.
 The parameters $\sin^2 2 \theta_{23}$ and $\Delta m^2_{32}$ should be readily 
accessible in all three experiments given the high event rates in the 
$\nu_\mu \to \nu_\mu$ channel with $\nu_\tau$ 
detection also possible in both ICARUS and OPERA.
Because of the expected smallness of $\sin^2 2 \theta_{13}$ and the $\nu_e$ 
background in the beam, $\nu_e$ appearance 
will be difficult to study, and the accessibility to $\sin^2 2 \theta_{13}$
 depends strongly on the detector. 

\section{MINOS Experiment}
\label{sec:minos}

The MINOS experiment~\cite{minos} is designed to detect 
neutrinos from the Fermilab NuMI beam.
The source of the neutrino beam is the decay of pions and kaons produced 
by collisions of 120 GeV protons (originating from the Fermilab Main Injector) 
with a nuclear target. 
A total of 3.7 $\times$ 10$^{20}$  protons on target are expected per year. The beam 
will be almost exclusively $\nu_\mu$ with a  
$(\nu_e+\bar \nu_e)$  contamination of about $1 \%$. 
A low, medium, or high energy neutrino beam can be realized at MINOS by  
adjusting a focusing horn at the source. The resulting beam energies 
are peaked at approximately $3$, $7$ and $15$ GeV, respectively.     
For MINOS to operate as a $\nu_\tau$ appearance experiment either the high or medium energy 
beam would be required, since the $\tau$ production threshold is 3.1 GeV. 
The experiment will most likely start with the low energy configuration, 
since this configuration will maximize its sensitivity to lower values 
of $\Delta m^2_{32}$, that are favored by the latest Super-Kamiokande 
and K2K results~\cite{fogli1}. We do not include the effects of a hadronic hose or beamplug
as these options are unlikely to be realized. 

There will be two iron/scintillator detectors associated with the MINOS beam. 
The \mbox{1 kton} near detector, which will 
detect neutrinos before oscillations occur, will be located 
on site at Fermilab. The 5.4 kton far detector will be located 732 km away from the source
in the Soudan mine in Minnesota. It is expected that the experiment will
 run for two to three years starting in 2005.
 
MINOS will independently measure the rates and the energy spectra for 
muonless ($0\mu$) and single-muon ($1\mu$) events, which are related to the neutral 
current (nc) and charged current (cc) reactions, respectively~\cite{minostec}.
The $0\mu$ and $1\mu$ event rates can
be used to measure the ratio of neutral current events to charged current events 
which is an important consistency check on neutrino oscillations and
could be useful for determining if a
sterile neutrino is involved in the disappearance channel. However, it does not
enhance the precision with which the oscillation parameters can be extracted. Since 
our goal is to determine the precision with which the oscillation parameters can be
found in a three neutrino framework, it is a sound assumption that $\nu_\mu$
disappearance is a consequence of transitions to $\nu_\tau$. We do not consider
$0\mu$ events in what follows.

We assume the low energy beam configuration 
to study the MINOS sensitivity
to $\Delta m^2_{32}$, $\sin^2 2\theta_{23}$ 
and $\sin^2 2\theta_{13}$. We focus on $\nu_\mu \to \nu_\mu$ disappearance 
to calculate the $1 \mu$-event energy spectrum of charged current $\nu_\mu$ 
events and $\nu_\mu \to \nu_e$ appearance to calculate the total integrated rate 
of charged current $\nu_e$ interactions. 

The $1 \mu$-event energy spectrum is divided into 23 bins of variable width  
$\Delta E_i$, according to the method described in Ref.~\cite{messier}. 
The $1 \mu$-event sample at MINOS will consist mainly of cc-events, with a 
small contribution coming from the misidentification of nc-events as cc-events. 
Thus, the content of the $i^{th}$ bin, $dN^{i}_{1 \mu}/dE$, is given by   
\begin{eqnarray}
\frac{dN^{i}_{1\mu}}{dE}(\Delta m^2_{32},\sin^2 2 \theta_{23},\sin^2
2\theta_{13}) = \frac{A}{\Delta E_i} 
\int_{E_i-\frac{\Delta E_i}{2}}^{E_i+\frac{\Delta E_i}{2}}  
\phi_{\nu_{\mu}}(E) & &   
\left[    
\sigma_{\nu_{\mu}}^{\text{cc}}(E) 
P_{\nu_\mu \rightarrow \nu_\mu}(E) \epsilon^{\text{cc}}(E) \right.
\nonumber \\
& &  + \left. 
\sigma_{\nu}^{\text{nc}}(E) \eta^{\text{nc}}(E)
\right] \, dE\,.
\label{minos-spec}
\end{eqnarray}
Here, $\phi_{\nu_{\mu}}$ is the neutrino flux at the MINOS far
detector.  We use the energy spectrum 
for $\nu_\mu$ cc-events in the MINOS far 
detector in Ref.~\cite{messier}.  The number of active targets is 
$A = M \, 10^9 \, N_{A} \, n_{p} \, n_y$, where $M$ 
is the detector mass in kton,  $10^9 \, N_{A}$ is the number of nucleons 
per kton ($N_{A}\equiv$ Avogadro's number),   
$n_y$ is the number of years of data taking and $n_{p}$ is the number of 
protons on target per year. The $\nu_\mu$ charged current cross section, 
$\sigma_{\nu_\mu}^{\text{cc}}$, and
the neutral current cross section, $\sigma_{\nu}^{\text{nc}}$,
are provided in Ref.\ \cite{phd} and Ref.\ \cite{nc}, respectively.  
Here, $\epsilon^{\text{cc}}$ is the probability that a given charged current 
$\nu_\mu$ event will be correctly labelled ($\sim$ 80\% at the spectral peak), and 
$\eta^{\text{nc}}$ is the probability of misidentifying a given neutral current 
event as a $1 \mu$ event ($\sim$ 10\% at the spectral peak). These factors were estimated
by the MINOS collaboration using a detailed Monte Carlo simulation of the detector with the 
low energy beam configuration~\cite{petyt}. The Standard
Model expectation at the far detector is about 400 single-muon
events per kton-year.

The MINOS experiment will reduce the systematic errors associated with
the charged current energy spectrum 
by comparing the spectrum measured at the far detector 
with that measured at the near detector from the same beam. 
The uncertainties in the far/near ratio, mainly due to the 
theoretical uncertainties in the secondary production of $\pi$ and $K$ in the 
NuMI target, range from 1-4\% in the peak of the low energy beam up to 
5-10\% in the high energy tail~\cite{messier}. 
We have neglected these errors in our study.  

The $\nu_e$ rate is defined as   
\begin{equation}
R_{e}(\Delta m^2_{32},\sin^2 2\theta_{23},\sin^2 2\theta_{13}) =
S_e+B^{({\mu}{\tau})}_{e}+ B^{({\mu}{\mu})}_{e}+
B^{(\text{beam})}_{e}+B^{\text{(nc)}}_{e}\,,
\label{mu-el-minos}
\end{equation}
where the signal, $S_{e}$, is 
\begin{eqnarray}
S_{e}(\Delta m^2_{32},\sin^2 2\theta_{23},\sin^2 2\theta_{13}) =
A \int   \phi_{\nu_{\mu}}(E) \, P_{\nu_{\mu} \to 
\nu_{e}}(E) \sigma_{\nu_{e}}^{\text{cc}}(E)\, \epsilon_e  \, dE\,.
\end{eqnarray}
Here, $\epsilon_e=0.28$ is the signal efficiency and $ \sigma_{\nu_{e}}^{\text{cc}}$ is 
the $\nu_e$ charged current cross section~\cite{phd}. 
The electron background event contribution from the decay of tau leptons in the
detector is   
\begin{eqnarray}
B^{({\mu}{\tau})}_{e}
(\Delta m^2_{32},\sin^2 2\theta_{23},\sin^2 2\theta_{13}) =
A \int   \phi_{\nu_{\mu}}(E) \, P_{\nu_{\mu} \to 
\nu_{\tau}}(E) \sigma_{\nu_{\tau}}^{\text{cc}}(E) \,
\epsilon^{({\mu}{\tau})}_{\text{R}} 
dE \,,
\end{eqnarray}
where $\epsilon^{({\mu}{\tau})}_{\text{R}}=0.066$ 
is the reduction efficiency and 
$\sigma_{\nu_{\tau}}^{\text{cc}}$ is the $\nu_\tau$ charged 
current cross section~\cite{phd}. The background from 
$e/\mu$ misidentification is given by  
\begin{eqnarray}
B^{({\mu}{\mu})}_{e}
(\Delta m^2_{32},\sin^2 2\theta_{23},\sin^2 2\theta_{13}) =
A \int   \phi_{\nu_{\mu}}(E) \, P_{\nu_{\mu} \to 
\nu_{\mu}}(E) \sigma_{\nu_{\mu}}^{\text{cc}}(E)
\eta^{({\mu}{\mu})} dE &,
\end{eqnarray}
where $\eta^{({\mu}{\mu})}=0.001$ is the misidentification
probability. The background coming from the $\nu_e$ beam contamination is 
\begin{eqnarray}
B^{(\text{beam})}_{e}
(\Delta m^2_{32},\sin^2 2\theta_{23},\sin^2 2\theta_{13}) =
A \int   \phi_{\nu_{e}}(E) \, P_{\nu_{e} \to 
\nu_{e}}(E) \sigma_{\nu_{e}}^{\text{cc}}(E) \, 
\epsilon^{(\text{beam})}_{\text{R}}
dE &,
\end{eqnarray}
where $\epsilon^{(\text{beam})}_{\text{R}}=0.079$    
is the reduction efficiency. The final background included, $B^{(\text{nc})}_{e}$,
from the decay of neutral pions created by neutral current interactions, 
is
\begin{eqnarray}
B^{(\text{nc})}_{e}
(\Delta m^2_{32},\sin^2 2\theta_{23},\sin^2 2\theta_{13}) =
A \int   \phi_{\nu_{\mu}}(E) \, \sigma_{\nu}^{\text{nc}}(E) 
\, \epsilon^{(\text{nc})}_{\text{R}} 
dE &,
\end{eqnarray}
where $\epsilon^{(\text{nc})}_{\text{R}}=0.016$ is 
the reduction efficiency. The efficiencies used in calculating both the 
signal and the backgrounds for MINOS $\nu_e$ events can be found in 
Ref.~\cite{diwan}.

\section{CERN-Gran Sasso Experiments}
\label{sec:opera}

A new facility under construction at CERN will 
direct a $\nu_\mu$ beam 732 km to the 
Gran Sasso Laboratory in Italy where it will be intercepted by two massive detectors, 
ICARUS and OPERA. 
The number of protons on target at the CERN SPS source is expected to be 4.5 $\times 
10^{19}$ per year, and the $\nu_\mu$ beam will have an average energy of 17 GeV. The  
fractions  $\nu_e/\nu_\mu$, $\nu_{\bar{\mu}}/\nu_\mu$  and 
$\nu_\tau/\nu_\mu$  in the beam are expected 
to be as low as 0.8\%, 2\% and $10^{-7}$, respectively. 

The
ICARUS~\cite{icarus} detector  
 will use liquid argon for its detection medium, and is expected
to have an initial effective volume of 3 ktons with a 10 year running time.
We conservatively assume an exposure 
of 20 kton-years because more data than this does not significantly improve either
the reach of the experiment or the precision with which the oscillation parameters are
determinable.  
 OPERA~\cite{opera}, another detector at the Gran Sasso consists of
lead plates interspaced with emulsion sheets, and is expected to have an
effective volume of 2 kton and a running time of 5 years.      

We investigate the capabilities of ICARUS and OPERA as $\nu_\mu$ disappearance, $\nu_e$ 
appearance, and $\nu_\tau$ appearance experiments. We calculate the full energy spectrum 
for the charged current $\nu_\mu$ events, and rates for the charged current $\nu_e$ and 
$\nu_\tau$ events. Low statistics in the $\nu_e$ and $\nu_\tau$ appearance channels makes a 
study of their full energy spectra unfeasable.    
The $\nu_\mu$ scattering energy spectra in ICARUS and OPERA consists of 
16 bins of width $\Delta E=2.5$ GeV. The number of events in the $i^{th}$ bin, 
$dN^{i}_{\mu}/dE$,  
is calculated as  
\begin{equation}
\frac{dN^{i}_{\mu}}{dE}(\Delta m^2_{32},\sin^2 2\theta_{23},\sin^2 2\theta_{13}) = \frac{A}{\Delta E} \int_{E_i-1.25 
\text{ GeV}}^{E_i+1.25 \text{ GeV}} \phi_{\nu_{\mu}}(E) \, 
P_{\nu_{\mu} \to \nu_{\mu}}(E) \, \sigma_{\nu_{\mu}}^{\text{cc}}(E) \, \epsilon(E) \; dE,
\label{numu}
\end{equation}
where $\phi_{\nu_{\mu}}$ is the flux of $\nu_\mu$ arriving at the Gran Sasso 
laboratory and  $\sigma_{\nu_{\mu}}^{\text{cc}}$ is the charged current 
cross section for $\nu_{\mu}$-nucleon scattering, both of which are available in 
Ref.~\cite{x-sec}.  
Again, $A$ is the number of active targets in the  detector, and we assume a constant 
efficiency $\epsilon=0.98~(0.94)$ for ICARUS (OPERA) according to 
Refs.\cite{icarus,icanoe} and~\cite{opera2}, respectively. If no oscillations occur, 
2180 (2070) $\nu_\mu$ charged current events per kton-year 
are expected at the ICARUS (OPERA) detectors. 

The $\nu_\tau$ production rate at ICARUS, $R_\tau$, is 
\begin{eqnarray}
R_{\tau}(\Delta m^2_{32},\sin^2 2\theta_{23},\sin^2 2\theta_{13}) =
 S_{\tau}+ 
 B^{({\mu}e)}_{\tau}+ B^{(\text{beam})}_{\tau},
\end{eqnarray}
where the signal, $S_{\tau}$, is defined as  
\begin{equation}
S_{\tau}(\Delta m^2_{32},\sin^2 2\theta_{23},\sin^2 2\theta_{13}) =
A \displaystyle \int \phi_{\nu_{\mu}}(E) \, P_{\nu_{\mu} \to 
\nu_{\tau}}(E) \sigma_{\nu_{\tau}}^{\text{cc}}(E) \, \epsilon_\tau \; dE\,,
\label{nutau}
\end{equation}
and we use an overall efficiency $\epsilon_\tau = 0.06$~\cite{icarus,icanoe}.
Here we have assumed that the $\tau$ 
identification at ICARUS will be made via the $\tau \to e$ leptonic decay
only~\cite{icarus2,icanoe,ica_2000}. The overall efficiency therefore includes
the branching ratio which is $\sim 18\%$ and the 
selection cuts efficiency.

Since tau events will be detected via an electron final state at ICARUS, $\nu_e$
backgrounds in the beam must be considered for this experiment. One source is 
 $\nu_\mu \to \nu_e$ oscillations, and is given by 
\begin{equation}
B^{({\mu}e)}_{\tau} 
(\Delta m^2_{32},\sin^2 2\theta_{23},\sin^2 2\theta_{13}) = 
A \displaystyle \int \phi_{\nu_{\mu}}(E) \, P_{\nu_{\mu} \to 
\nu_{e}}(E) \sigma_{\nu_{e}}^{\text{cc}}(E) \, 
{\epsilon}^{({\mu}e)}_{\text{R}}\; dE\,.
\end{equation}
Another important background contribution comes from 
the intrinsic $\nu_e$ beam contamination, and is given by  
\begin{equation}
B^{(\text{beam})}_{\tau} 
(\Delta m^2_{32},\sin^2 2\theta_{23},\sin^2 2\theta_{13}) = 
A \displaystyle \int \phi_{\nu_{e}}(E) \, P_{\nu_{e} \to 
\nu_{e}}(E) \sigma_{\nu_{e}}^{\text{cc}}(E) \, 
{\epsilon}^{(\text{beam})}_{\text{R}}\; dE\,.
\end{equation}
The reduction efficiencies are ${\epsilon}^{({\mu}e)}_{\text{R}}=0.15$ and 
${\epsilon}^{(\text{beam})}_{\text{R}}=0.009$~\cite{icanoe,ica_2000}.  

For the OPERA experiment, we have assumed $\tau$ detection via its one-prong decay into 
leptons ($l$) and hadrons ($h$)~\cite{opera2}. Backgrounds can be safely neglected,
and the integrated rate of charged current $\nu_\tau$ interactions at OPERA is given by 
Eq.~(\ref{nutau}) with $\epsilon_\tau = 0.087$. This efficiency value includes 
$\sum_{i=l,h} \text{Br}(\tau \to i)$ as well as the 
selection cuts efficiency~\cite{opera2}.

The expression for the $\nu_e$ production rate at ICARUS and OPERA, $R_e$, 
is identical to the one presented in Eq.~(\ref{mu-el-minos}) with the 
appropriate fluxes, cross sections, and efficiencies. 
For ICARUS, $\epsilon_e=0.75$, 
$\epsilon^{({\mu}{\tau})}_{\text{R}}=0.14$, $\eta^{({\mu}{\mu})}=0.002$,
$\epsilon^{(\text{beam})}_{\text{R}}=0.19$  and 
$\epsilon^{(\text{nc})}_{\text{R}}=0.01$ so as to emulate the effect of the
selection cuts of Ref~\cite{icarus2}.
The corresponding parameters for OPERA are 
$\epsilon_e=0.7$, $\epsilon^{({\mu}{\tau})}_{\text{R}}=0.13$ 
(which reproduces the electron detection 
efficiency of Ref.~\cite{opera}),   
$\eta^{({\mu}{\mu})}=0.002$,
$\epsilon^{(\text{beam})}_{\text{R}}=0.19$  and 
$\epsilon^{(\text{nc})}_{\text{R}}=0.016$. 

\section{Data Simulation and Analysis}
\label{sec:ana}
In this section we describe the procedures used to determine the sensitivity and 
contour plots for the three experiments and their combination.

\subsection{$\chi^2$ DEFINITIONS}

\subsubsection{$\nu_\mu \rightarrow \nu_\mu$} 
\label{nu_mu}

For this oscillation channel we use the information from the 
$\nu_\mu$ charged current energy spectrum of each experiment. 
All three experiments will observe large numbers of events of this type, so that 
the number of events in each bin obeys Gaussian statistics. The $\chi^2$ function
is then defined as 
\beq
\chi^2_{\mu} (\Delta m^2_{32},\sin^2 2\theta_{23},\sin^2 2\theta_{13}) 
= \sum_{i,j=1,n_\beta} 
( N^{\text{obs}}_i-N^{\text{th}}_i) 
\sigma_{ij}^{-2} 
( N^{\text{obs}}_j-N^{\text{th}}_j), 
\label{chi2-mu}
\ee 
where $N^{\text{th}}_{i}$ denotes the theoretical prediction for the number of events
in bin $i$ (for a given set of oscillation parameters), calculated according to 
Eq.~(\ref{minos-spec}) for 
MINOS and Eq.~(\ref{numu}) for ICARUS and OPERA. $N^{\text{obs}}_i$ denotes the 
"observed" number of events in bin $i$. The simulation of the data giving 
$N^{\text{obs}}_i$ will be described in Section C.   
The number of bins is $n_{\beta}=$ 23, 16 and 16 for $\beta=$ MINOS, ICARUS and 
OPERA, respectively. For MINOS, the error matrix is 
defined as $\sigma_{ij}^2 = \delta_{ij}
N^{\text{obs}}_i +  (0.02)^2 N^{\text{obs}}_i N^{\text{obs}}_j$, where
the off-diagonal elements reflect a 2\% overall flux
uncertainty~\cite{arroyo}. For ICARUS and OPERA we only consider 
statistical errors, yielding $\sigma_{ij}^2 = \delta_{ij} N^{\text{obs}}_i$.

\subsubsection{$\nu_\mu \rightarrow \nu_e$}

Gaussian statistics will also be realized in this channel. Even in the 
case of no oscillations the background contributes a sufficient number of
events to $R_e$ to warrant a Gaussian $\chi^2$ function, 
\beq 
\chi^2_{e}(\Delta m^2_{32},\sin^2 2\theta_{23},\sin^2 2\theta_{13}) ={\left(
\frac{R^{\text{obs}}_e-
R^{\text{th}}_e}
{\sigma_{e}}\right)}^2\,. 
\label{chi2-el}
\ee
Here, the theoretical electron rate  $R^{\text{th}}_e$ is given
by Eq.~(\ref{mu-el-minos}) with the appropriate experimental parameters 
(fluxes, efficiencies, etc) for the experiment under consideration. For 
MINOS, we define the error 
as $\sigma^2_{e}=R^{\text{obs}}_e+
(\delta^{\text{min}}_{\text{syst}} \times R^{\text{obs}}_e)^2$ 
assuming a global systematic error 
$\delta^{\text{min}}_{\text{syst}}=0.1$. For ICARUS, we define 
the error as $\sigma^2_{e}=R^{\text{obs}}_e+(\delta^{\text{ica}}_{\text{syst}} 
\times  B^{(\text{beam})}_{e})^2$, where only the $\nu_e$ beam contamination has
been considered as a source of significant uncertainty with
$\delta^{\text{ica}}_{\text{syst}}=0.05$.   For OPERA, 
we use $\sigma^2_{e}=R^{\text{obs}}_e+
(\delta^{\text{ope}}_{\text{syst-I}} \times  B^{(\text{beam})}_{e})^2+
(\delta^{\text{ope}}_{\text{syst-II}}\times B^{\text{(nc)}}_{e})^2 $, 
where $B^{(\text{beam})}_{e}$ and $B^{\text{(nc)}}_{e}$ are the 
OPERA backgrounds produced by the $\nu_e$ beam contamination
and the misidentified neutral current interactions, respectively.
Their associated systematics errors are given by
$\delta^{\text{ope}}_{\text{syst-I}}=0.1$ 
and $\delta^{\text{ope}}_{\text{syst-II}}=0.2$, as suggested 
in Ref.~\cite{opera}. 
We do not include a systematic error for the misidentified neutral
current interactions at ICARUS because it has not been estimated.
 We neglect the errors arising from 
other components of the background because they have not been determined
on account of their expected insignificance. 
 
\subsubsection{$\nu_\mu \rightarrow \nu_\tau$}

The number of detected events in this channel is likely to be small for both ICARUS
and OPERA and so we use a $\chi^2$ function consistent with a 
Poisson distribution,
\beq 
\chi^2_{\tau}(\Delta m^2_{32},\sin^2 2\theta_{23},\sin^2 2\theta_{13}) =2\left[
\left(R^{\text{obs}}_{\tau}
-R^{\text{th}}_{\tau}\right)
+R^{\text{obs}}_{\tau} \ln{\left(
\frac{R^{\text{obs}}_{\tau}}
{R^{\text{th}}_{\tau}}\right)}
\right].
\label{chi2-tau} 
\ee
Here, the theoretical $\nu_\tau$ rate $R^{\text{th}}_{\tau}$ is calculated via
Eq.~(\ref{nutau}), using the appropriate experimental parameters for
ICARUS and OPERA. When the number of $\nu_\tau$ event rates is greater than 5, we use a standard
Gaussian $\chi^2$ function given by 
\beq 
\chi^2_{\tau}(\Delta m^2_{32},\sin^2 2\theta_{23},\sin^2 2\theta_{13}) ={\left(
\frac{R^{\text{obs}}_{\tau}
-R^{\text{th}}_{\tau}}
{\sigma_{\tau}}\right)}^2. 
\label{chi2-tau2}
\ee
The errors considered in this case are purely statistical.

\subsection{REGIONS OF SENSITIVITY}
\label{subsec:sensi}

For the $\nu_\mu \rightarrow \nu_e$ channel, we calculate
the sensitivity regions for MINOS, ICARUS and 
OPERA in the $(\sin^2 2\theta_{13},\Delta m^2_{32})$ plane. 
The procedure used is to set  
$R^{\text{obs}}_e=R_{e}(\Delta m^2_{32},\sin^2 2\theta_{23},0)$
with $\sin^2 2\theta_{23}=0.85${\footnote{This value is approximately the smallest
value of $\sin^2 2\theta_{23}$ allowed at the 99\% C. L. by a combined analysis of
SuperKamiokande and K2K data~\cite{fogli1}. 
We conservatively assume that $\theta_{23}<\pi/4$. If the converse were true, the sensitivity
to $\sin^2 2\theta_{13}$ is greater than for $\sin^2 2\theta_{23}=1$.}} or $1$, and  
minimize the $\chi^2$ function given in Eq.~(\ref{chi2-el}) 
with respect to the parameters $\sin^2 2\theta_{13}$ and 
$\Delta m^2_{32}$. Curves are then drawn at $\Delta \chi^2 = \chi^2 - 
\chi^2_{\text{min}} = 4.61$ corresponding to the $90\%$ C.L.. 
The sensitivities are calculated for each experiment separately as
well as for their combined potential by using the same procedure with
$\chi^2=\chi^2_{e(\text{minos})}+\chi^2_{e(\text{icarus})}+\chi^2_{e(\text{opera})}$. 

We determine the sensitivity to $\sin^2 2\theta_{23}$ and $\Delta m^2_{32}$ in the 
$\nu_\mu \rightarrow \nu_\mu$ channel for MINOS and in the 
$\nu_\mu \rightarrow \nu_\tau$
channel for ICARUS and OPERA. For MINOS, we calculate  
$ N^{\text{obs}}_i= dN^{i}_{1\mu}/dE(\Delta m^2_{32},0,\sin^2 2\theta_{13})$
with $\sin^2 2\theta_{13}$ set either to 0 or 0.1, its minimum or maximum allowed value. 
Equation~(\ref{chi2-mu}) is then minimized
with respect to $\sin^2 2\theta_{23}$ and $\Delta m^2_{32}$, and the contour of
constant $\Delta \chi^2$ at the $90\%$ C.L. is found. 
For ICARUS and OPERA, we set
$R^{\text{obs}}_{\tau}=R_{\tau}(\Delta m^2_{32},0,\sin^2 2\theta_{13})$ in 
Eq.~(\ref{chi2-tau})/(\ref{chi2-tau2}),
again with $ \sin^2 2\theta_{13}=0$ or 0.1, minimize with respect to
 $\sin^2 2\theta_{23}$ and $\Delta m^2_{32}$ and trace contours of constant
$\Delta \chi^2$. 
For the combined sensitivity of all three experiments, 
the same procedure is repeated with a global $\chi^2$ 
defined as  $\chi^2=\chi^2_{\mu(\text{minos})}+
\chi^2_{\tau(\text{icarus})}+\chi^2_{\tau(\text{opera})}$. 

\subsection{PREDICTING THE ALLOWED REGIONS}

To predict how accurately MINOS, ICARUS and OPERA will determine
the neutrino 
oscillation parameters at the atmospheric scale, we simulate data sets for 
the three experiments assuming oscillations occur with  
\bea 
 \Delta m^2_{32} &=& 3 \times 10^{-3}\, {\rm{eV}}^2\,, \ \ \   
 \Delta m^2_{21} = 5 \times 10^{-5}\, {\rm{eV}}^2\,, \nonumber \\ 
\sin^2 2 \theta_{23} &=&  1\,, \ \ \    
\sin^2 2 \theta_{12} = 0.8\,, \ \ \ 
\sin^2 2 \theta_{13} = 0.05.
\label{input}
\eea 
All parameters other than $\theta_{13}$ are close to 
the best-fit values from the current data. 
The value of $\theta_{13}$ is chosen to lie within (or close to) the 
sensitivity regions of all three experiments under consideration while obeying the
CHOOZ limit~\cite{chooz}.
We have assumed a normal hierarchy ($\Delta m^2_{32} > 0$). Note that since matter
effects are small at 730 km, the results for an inverted hierarchy will not
be substantially different from that for a normal hierarchy.

Each simulated data point is generated by randomly choosing a point from a Gaussian 
 or Poisson distribution associated with the number of events expected theoretically. 
This procedure is followed 
for the 23 bins of the MINOS $\nu_\mu$ energy spectrum as well as
the 16 bins of the ICARUS and OPERA $\nu_\mu$ spectra, giving the $N^{\text{obs}}_i$
of Eq.~(\ref{chi2-mu}). This is also done for the $\nu_e$ charged current rates of all three
experiments, yielding the $R^{\text{obs}}_e$ in Eq.~(\ref{chi2-el}), and the $\nu_\tau$
charged current rates at ICARUS and OPERA, giving the $R^{\text{obs}}_\tau$ of 
Eq.~(\ref{chi2-tau}) or (\ref{chi2-tau2}). 
In the case of the $\nu_e$ and $\nu_\tau$ rates where backgrounds are 
taken into consideration, the backgrounds and signals are calculated and simulated 
separately before being combined to give the rates used in the $\chi^2$ analyses.
 
The allowed regions are calculated for the individual experiments
by combining the information available from each channel and defining total 
$\chi^2$ functions as follows:
\bea
\chi^2_{\text{minos}}(\Delta m^2_{32},\sin^2 2\theta_{23},\sin^2 2\theta_{13}) &=& 
\chi^2_{\mu}+\chi^2_{e}\,, \nonumber \\
\chi^2_{\text{icarus}}(\Delta m^2_{32},\sin^2 2\theta_{23},\sin^2 2\theta_{13}) &=& 
\chi^2_{\mu}+\chi^2_{e}+\chi^2_{\tau}\,, \nonumber \\
\chi^2_{\text{opera}}(\Delta m^2_{32},\sin^2 2\theta_{23},\sin^2 2\theta_{13}) &=& 
\chi^2_{\mu}+\chi^2_{e}+\chi^2_{\tau}\,.
\label{chi2-tot}
\eea
The $\chi^2$ functions are then minimized by varying $\Delta m^2_{32}$, $\sin^2
2\theta_{23}$ and $\sin^2 2\theta_{13}$ with
$\Delta m^2_{21}= 5 \times 10^{-5}$ eV$^2$ and
$\sin^2 2 \theta_{12} = 0.8$. Contours of $\Delta \chi^2$ = 6.25 and 11.36 are 
made corresponding to the $90\%$ and $99\%$ C.L., respectively. These confidence level
regions are projected on to two-dimensional subspaces of the three parameter space. 

To investigate whether combining data from the three experiments improves 
the determination of the oscillation parameters, we define a global $\chi^2$ 
function as 
\beq
\chi^2_{\text{minos+icarus+opera}}(\Delta m^2_{32},\sin^2
2\theta_{23},\sin^2 2\theta_{13}) = \chi^2_{\text{minos}} 
+ \chi^2_{\text{icarus}} + \chi^2_{\text{opera}}. 
\label{chi2-tot-all}
\ee
This $\chi^2$ function is minimized and contours are made as described above. 
 
\section{results}
\label{sec:res}
\subsection{MINOS}
\label{subsec:minos}

\begin{figure}
\vglue -1.0cm
\hglue -0.5cm
\centerline{
\epsfig{file=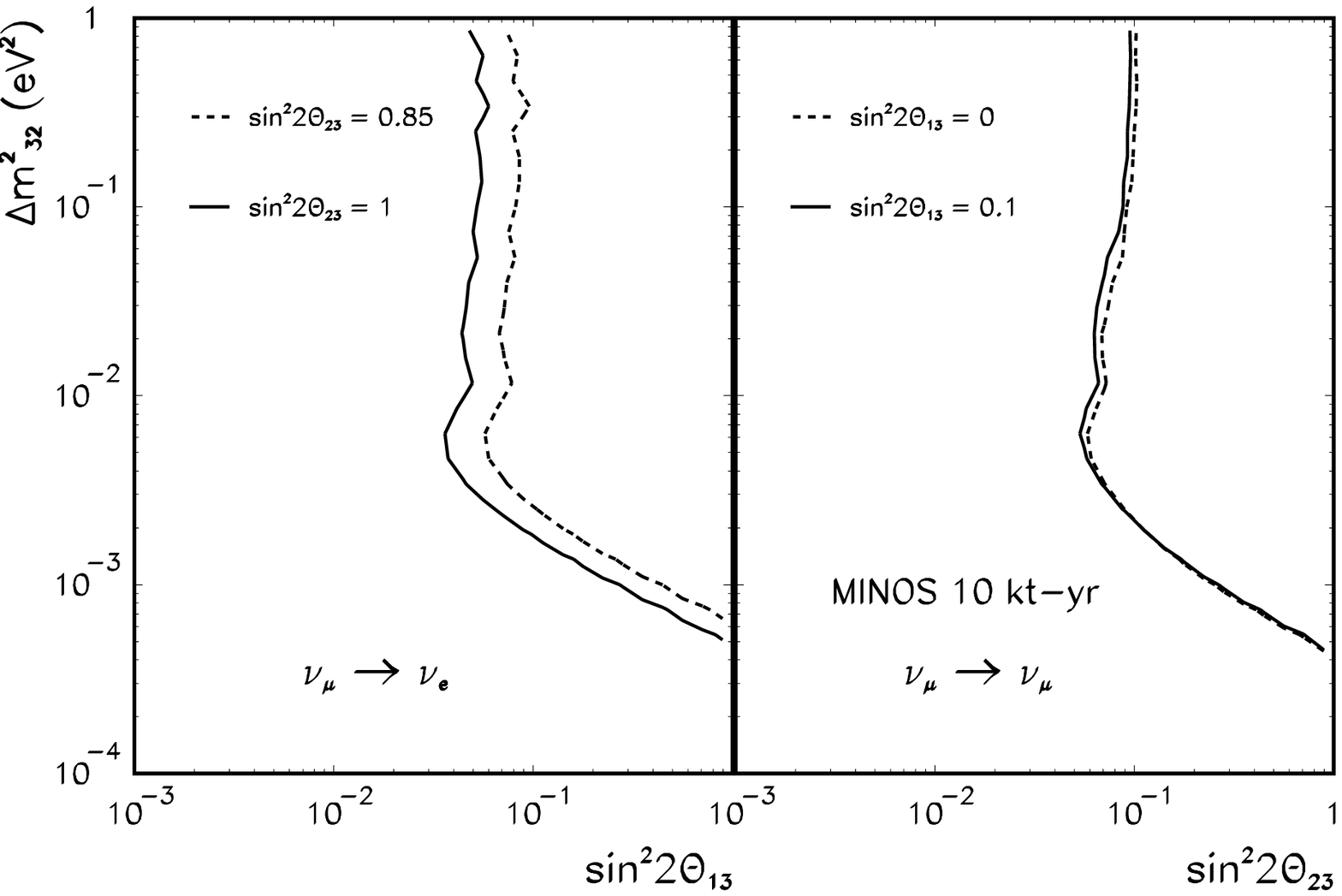,height=14.82cm,width=16.cm}}
\vglue -0.5cm
\caption[]{MINOS sensitivities for 
the $\nu_\mu \to \nu_e$ and $\nu_\mu \to \nu_\mu$ channels at the 90\% C.L..
The other oscillation parameters are fixed 
at $\Delta m^2_{21}=5 \times 10^{-5}$ eV$^2$ and 
$\sin^2 2\theta_{12}=0.8$, the best-fit solution to the solar neutrino problem.
To determine the sensitivity to $\sin^2 2\theta_{13}$ when $\sin^2 2\theta_{23}=0.85$,
we have taken $\theta_{23}<\pi/4$. The dashed curve would be to the left of the 
solid curve in the left panel if $\theta_{23}>\pi/4$.
}
\label{fig1a}
\vglue -1.cm
\end{figure}

In Fig.~\ref{fig1a} we present 
the sensitivity (at the 90\% C.L.) of MINOS in the $\nu_e$ appearance and 
$\nu_\mu$ disappearance 
 channels for an exposure of 10 kton-years. 
In the $\nu_e$ appearance channel (left panel of Fig.~\ref{fig1a}), 
we see a small decrease in the 
sensitivity of MINOS as $\sin^2 2\theta_{23}$ is decreased within its
allowed range. MINOS is sensitive to
 $\sin^2 2\theta_{13} \gsim 0.05$ for \mbox{$\Delta m^2_{32} \sim 3 \times 10^{-3}$ eV$^2$} 
and maximal $\nu_\mu-\nu_\tau$ mixing.
For the $\nu_\mu$
disappearance channel, one can see that the region of
sensitivity is not affected by the variation of $\sin^2 2 \theta_{13}$
within its allowed range.
MINOS is sensitive to $\Delta m^2_{32}\gsim 5 \times 10^{-4}$ eV$^2$ 
at maximal $\sin^2 2\theta_{23}$.
The sensitivity of the neutral current to charged current event ratio 
to the leading oscillation parameters has been analysed in Ref.~\cite{petyt}.
That sensitivity is comparable to the sensitivity of the $\nu_\mu$ disappearance channel
in Fig.~\ref{fig1a}
Direct evidence
for transitions to $\nu_\tau$ is a very important aspect of the MINOS experiment, 
which can be accomplished by comparing the $0\mu$ and $1\mu$ event rates. 

For the oscillation parameters of Eq.~(\ref{input}), the signal (background) is comprised of 
2770 (0) 
$\nu_\mu$ cc events and 15 (41) $\nu_e$ events. Note that we have included the 
contribution from misidentified nc-events in the $\nu_\mu$ event sample.
In Fig.~\ref{fig1b} we show the 
allowed regions obtained by simulating data for the MINOS detector.
The $\Delta\chi^2$ contours are shown at the 90 and 99\% C.L..
Because of high statistics
and an optimal L/E combination, MINOS should be able to pin down 
$\Delta m^2_{32}$ and $\sin^2 2\theta_{23}$ quite precisely, as is shown
most clearly by the plot in the $(\sin^2 2\theta_{23}, \Delta m^2_{32})$ 
plane. For 10 kton-years of exposure, 
we find that MINOS should be able to measure these parameters 
to within 10\% at the 99\% C.L.. However, MINOS will not 
make a determination of $\sin^2 2\theta_{13}$. Low 
statistics in the $\nu_e$ channel are expected because of the smallness of 
$\theta_{13}$, and relatively large backgrounds of $\nu_e$ are expected to be present 
in the beam. 

\subsection{ICARUS}
\label{subsec:icarus}

The sensitivity of the ICARUS experiment to the 
$\nu_\mu \to \nu_e$ and $\nu_\mu \to \nu_\tau$ oscillation channels are shown in 
the left and right-hand panels of Fig.~\ref{fig3a}, respectively.
20 kton-years of exposure is assumed, and the contours are 
drawn at the 90 and 99\% C.L.. 
For the $\nu_e$ appearance channel, we find that ICARUS can access values of 
$\sin^2 2\theta_{13} \gsim 0.03$ for 
$\Delta m^2_{32} \sim 3 \times 10^{-3}$ eV$^2$ and maximal $\nu_\mu-\nu_\tau$ mixing.
From the $\nu_\tau$ appearance channel, we find that 
ICARUS is sensitive to $\Delta m^2_{32}\gsim 1-2 \times 10^{-3}$ eV$^2$ at maximal
mixing.

The theoretical input (Eq.~(\ref{input})) yields 41960 (0), 34 (380) and 27 (11) 
$\nu_\mu$, $\nu_e$, 
$\nu_\tau$ signal (background) events, respectively.  
Figure~\ref{fig3b} shows 90 and 99\% C.L. contours obtained by simulating data
for all three channels that can be studied by the ICARUS experiment by
observing variations in $\chi^2=\chi^2_e+\chi^2_\mu+\chi^2_\tau$. From the 
three panels displayed in the figure, it can be seen that ICARUS can determine 
$\Delta m^2_{32}$ and $\sin^2 2\theta_{23}$ to within 30\% at the 99\% C.L. but will
not provide a meaningful determination of $\sin^2 2\theta_{13}$.

\subsection{OPERA}
\label{subsec:opera}

Figure~\ref{fig2a} shows the
sensitivity of the OPERA experiment to oscillation parameters at the atmospheric scale via the 
$\nu_\mu \to \nu_e$ and $\nu_\mu \to \nu_\tau$ channels, respectively. 
This plot is made for 10 kton-years 
of exposure, and represents the 90\% C.L..
Using $\nu_e$ appearance, we find that 
OPERA will be sensitive to values of 
$\sin^2 2\theta_{13} \gsim 0.2$ for 
$\Delta m^2_{32} \sim 3 \times 10^{-3}$ eV$^2$ and maximal $\nu_\mu-\nu_\tau$ mixing, 
but this region is ruled out by CHOOZ. 
 Exploiting $\nu_\tau$ appearance,
OPERA is shown to be sensitive to $\Delta m^2_{32}\gsim 1\times 10^{-3}$ 
eV$^2$ at maximal mixing independent of the value of $\sin^2 2\theta_{13}$. 
Although OPERA has no considerable background for $\tau$ 
events as well as a higher 
detection efficiency than ICARUS, the latter has a greater sensitivity in this channel
because of its larger volume and consequently
higher event rate. 

The number of $\nu_\mu$, $\nu_e$, $\nu_\tau$ signal (background) events expected for the
oscillation parameters used are 19870 (0), 14 (230) and 16 (0),
respectively.  
In Fig.~\ref{fig2b} we present contours of $\Delta\chi^2$ at the 90 and
99\% C.L. for OPERA.

\subsection{Global Analysis}
\label{subsec:glo-comb}

In Fig.~\ref{fig4a} we present the 90\% C.L sensitivity regions after combining the
data simulated for the MINOS, ICARUS and OPERA experiments.
In the $(\sin^2 2\theta_{13},\Delta m^2_{32})$ plane, the combined sensitivity
does not show much improvement over ICARUS alone, resulting in a sensitivity
to $\sin^2 2\theta_{13} \gsim 0.02$ for 
$\Delta m^2_{32} \sim 3 \times 10^{-3}$ eV$^2$ and maximal $\nu_\mu-\nu_\tau$ mixing.
 The sensitivity of the experiments in the $(\sin^2 2\theta_{23},\Delta m^2_{32})$ 
plane depends only slightly on the value of $\sin^2 2\theta_{13}$ between
$0$ and $0.1$. The combined sensitivity of the three experiments
 is $\Delta m^2_{32}\gsim 4-5 \times 
10^{-4}$ eV$^2$ at maximal mixing which is the same as that 
of MINOS. 

In Fig.~\ref{fig4b} we show the allowed regions obtained from the 
combined analysis at the 90\%, 95\% and 99\% C.L..
The precision with which the combination of the experiments can determine
$\Delta m^2_{32}$ and $\sin^2 2\theta_{23}$ does not improve from 
the analysis with MINOS alone, since changes in $\chi^2_{\text{minos}}$ 
dominates over any changes in the overall $\chi^2$ function. 
These parameters can therefore
be determined to within 10\% once again. However, the ability of these
experiments to constrain $\sin^2 2\theta_{13}$ becomes feasible
by combining them into a single analysis. For the data we simulated, we 
obtain $\sin^2 2\theta_{13} \gsim 0.01$ at the 95\% C.L.. Similar 
results for $\sin^2 2\theta_{13}$ are obtained by setting $\chi^2 =  
\chi^2_{\text{minos}} + \chi^2_{\text{icarus}}$, since only MINOS and ICARUS 
are sensitive to values of $\sin^2 2\theta_{13}<0.1$
 for $\Delta m^2_{32}=3 \times 10^{-3}$ eV$^2$. 
               
\section{Conclusions}
\label{sec:conc}

We have studied the potential of the future long baseline experiments
ICARUS, MINOS, and OPERA, to ascertain the neutrino oscillation parameters
at the atmospheric scale as separate experiments, and in combination. 
By simulating data at 
$\Delta m^2_{32} = 3 \times 10^{-3}$ eV$^2$, 
$\Delta m^2_{21} = 5 \times 10^{-5}$ eV$^2$, 
$\sin^2 2 \theta_{23} =  1$, $\sin^2 2 \theta_{12} = 0.8$ and 
$\sin^2 2 \theta_{13} = 0.05$ for all three experiments, and
calculating the allowed regions given by this data, we have
estimated the precision with which these
experiments can determine the parameters $\Delta m^2_{32}$, 
$\sin^2 2\theta_{23}$, and $\sin^2 2\theta_{13}$. We took
 into consideration $\nu_\mu$ disappearance and $\nu_e$ appearance
for all three experiments and $\nu_\tau$
appearance for ICARUS and OPERA. The precision with which the leading 
oscillation parameters can be determined by the combined data of the 
three experiments is illustrated in Fig.~\ref{fig4b}.

\begin{figure}
\vglue -1.5cm
\hglue -0.5cm
\centerline{
\epsfig{file=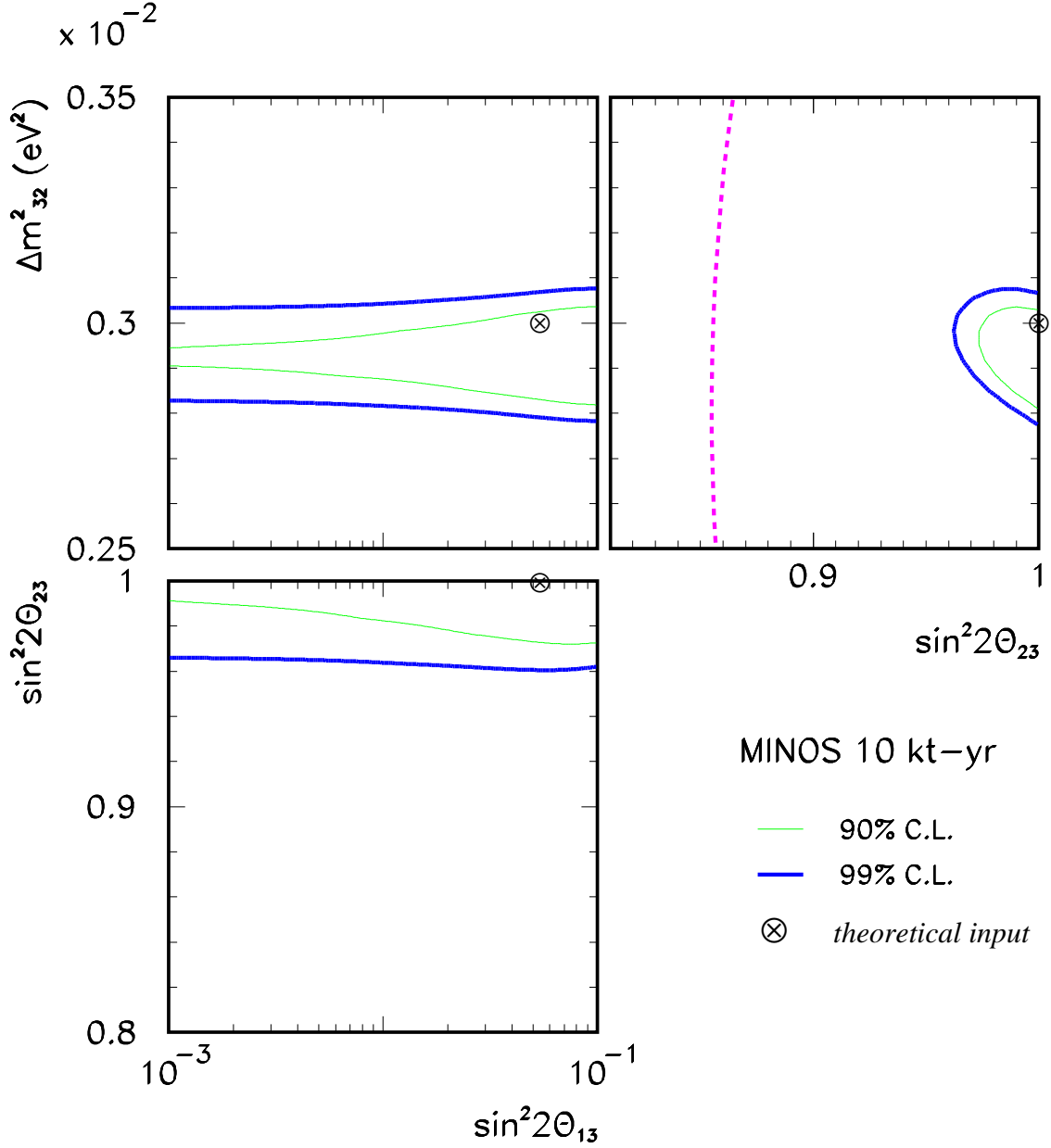,height=25cm,width=18.5cm}}
\vglue -4.5cm
\caption{Expected allowed regions for MINOS at the 90\% and 99\% 
C.L. using
$\Delta m^2_{32}= 3 \times 10^{-3}$ eV$^2$,  
$\Delta m^2_{21}= 5 \times 10^{-5}$ eV$^2$, 
$\sin^2 2 \theta_{23} =  1$,   
$\sin^2 2 \theta_{12} = 0.8$ and
$\sin^2 2 \theta_{13} = 0.05$ as the theoretical input for
which data was simulated.
The dashed line is the Super-Kamiokande 
allowed region at the 99\% C.L.. Here and in other figures, the best fit point is 
very close to the theoretical input and is consequently not shown.
}
\label{fig1b}
\vglue -1.cm
\end{figure}

\begin{figure}
\vglue -1.0cm
\hglue -0.5cm
\centerline{
\epsfig{file=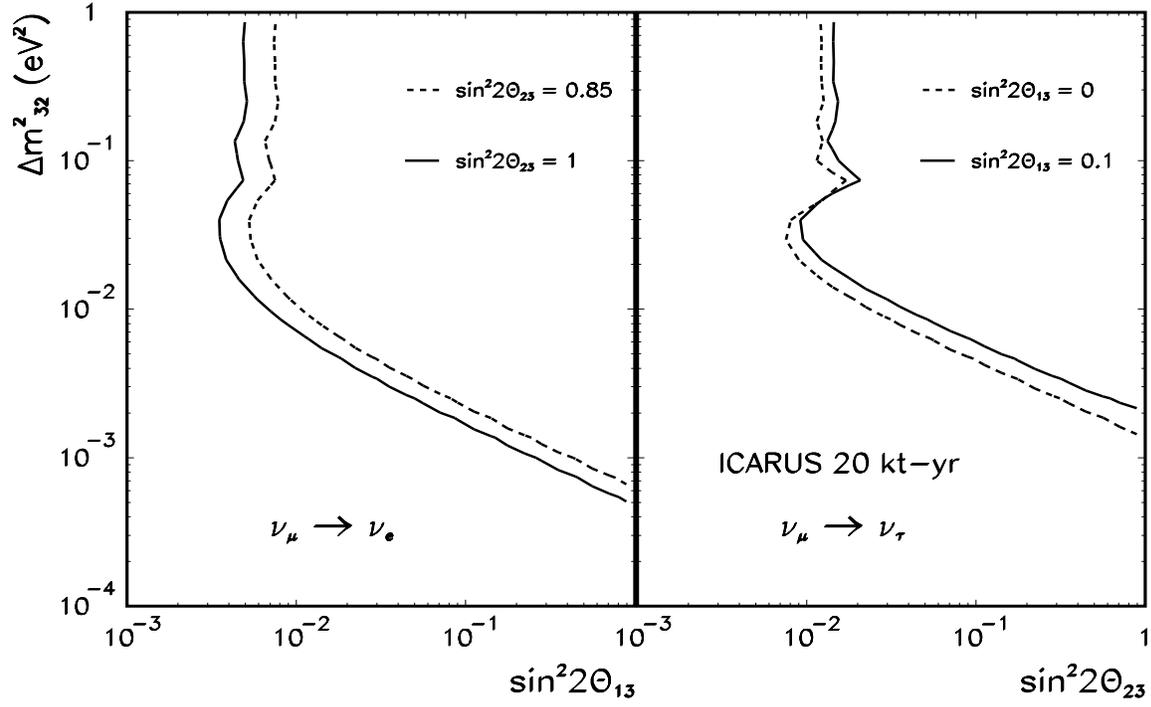,height=14.82cm,width=16.cm}}
\vglue -0.5cm
\caption{The same as Fig.~\ref{fig1a} but for ICARUS in 
the $\nu_\mu \to \nu_e \, (\nu_\tau)$ oscillation modes.} 
\label{fig3a}
\vglue -1.cm
\end{figure}

\begin{figure}
\vglue -1.5cm
\hglue -0.5cm
\centerline{
\epsfig{file=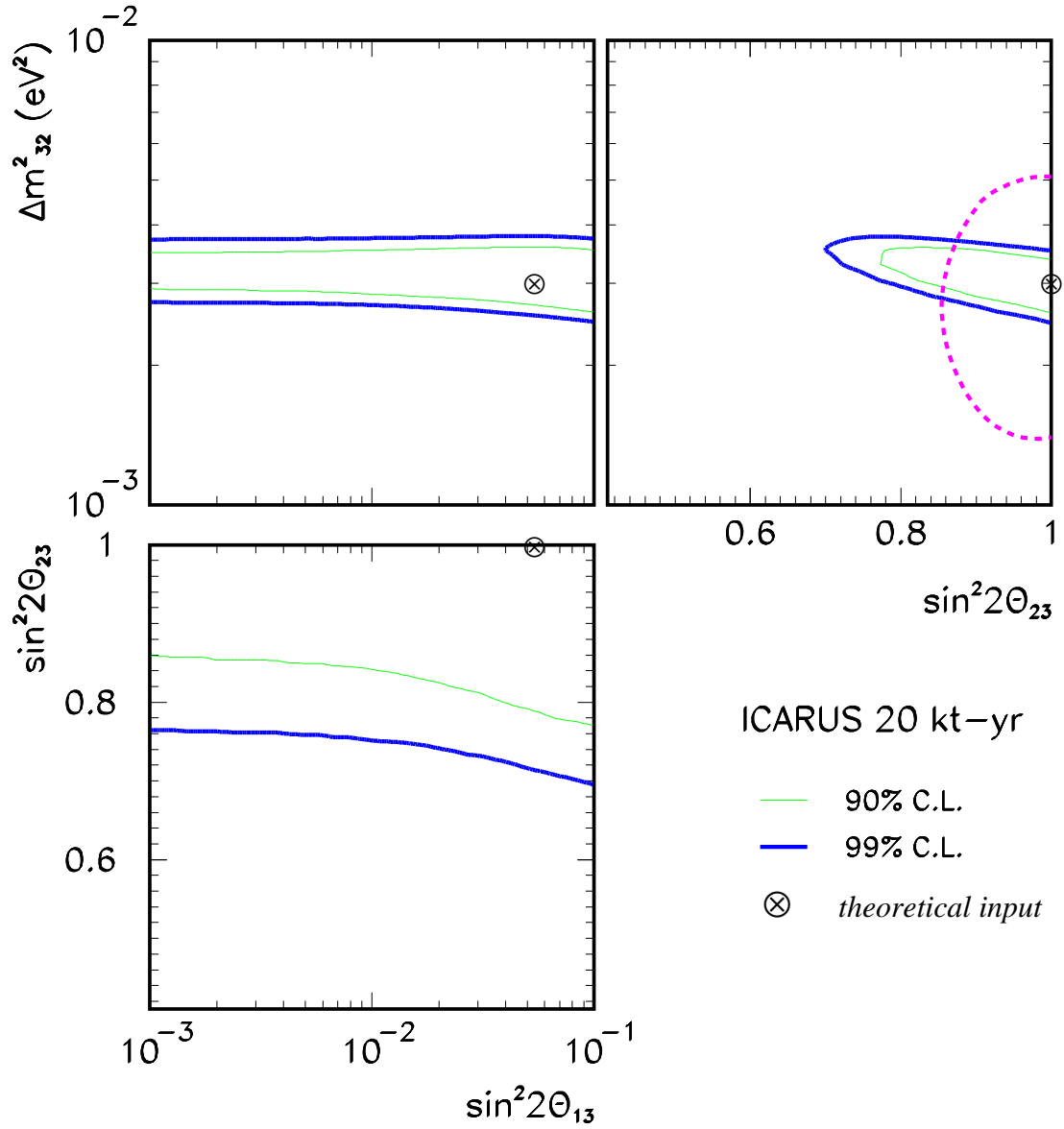,height=25.0cm,width=18.5cm}}
\vglue -4.5cm
\caption{The same as Fig.~\ref{fig1b} but for ICARUS.}
\label{fig3b}
\vglue -1.cm
\end{figure}

\begin{figure}
\vglue -1.0cm
\hglue -0.5cm
\centerline{
\epsfig{file=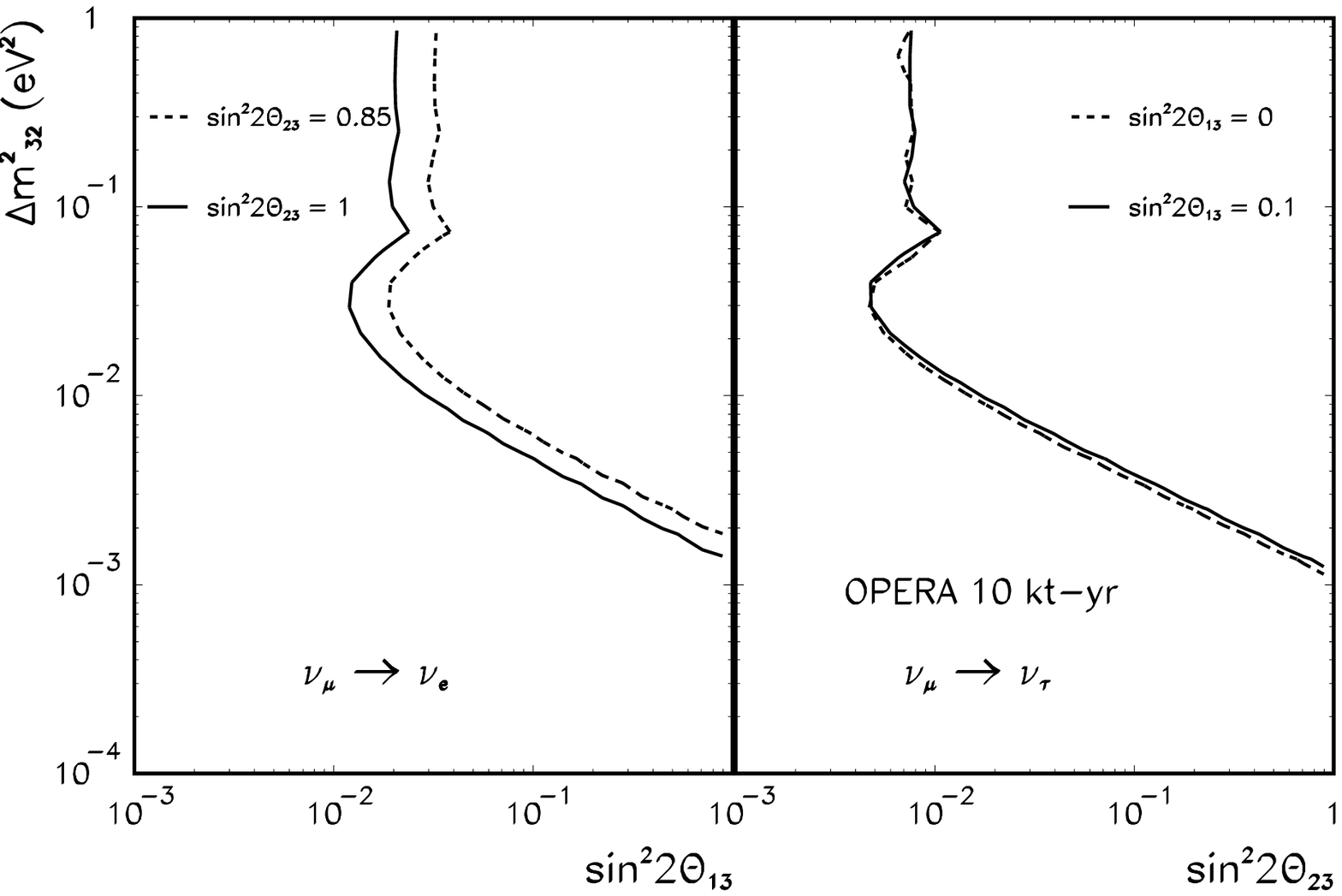,height=14.82cm,width=16.cm}}
\vglue -0.5cm
\caption{The same as Fig.~\ref{fig3a} but for OPERA.}
\label{fig2a}
\vglue -1.cm
\end{figure}

\begin{figure}
\vglue -1.5cm
\hglue -0.5cm
\centerline{
\epsfig{file=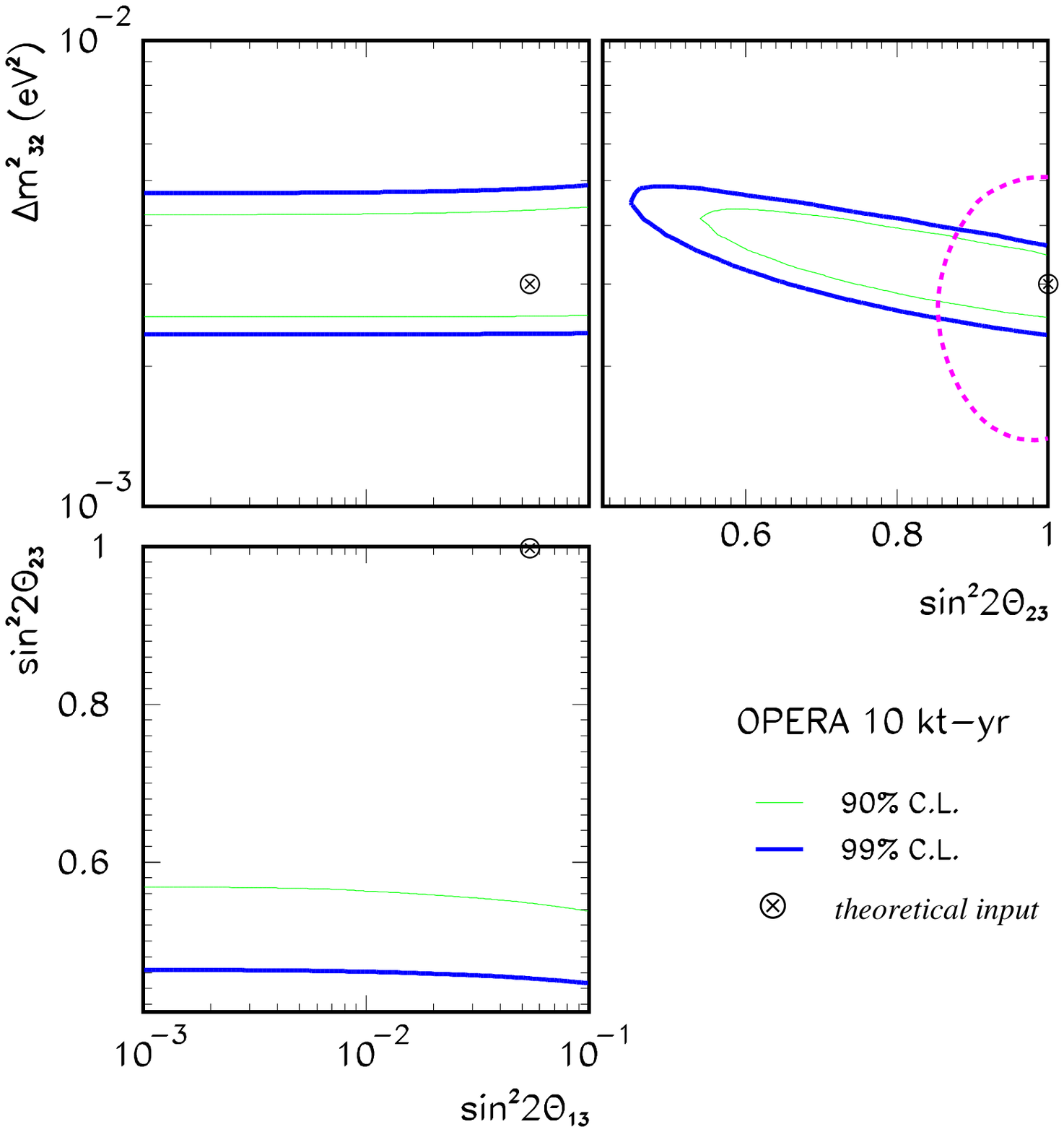,height=25.0cm,width=18.5cm}}
\vglue -4.5cm
\caption{The same as Fig.~\ref{fig1b} but for OPERA.}
\label{fig2b}
\vglue -1.cm
\end{figure}

\begin{figure}
\vglue -1.0cm
\hglue -0.5cm
\centerline{
\epsfig{file=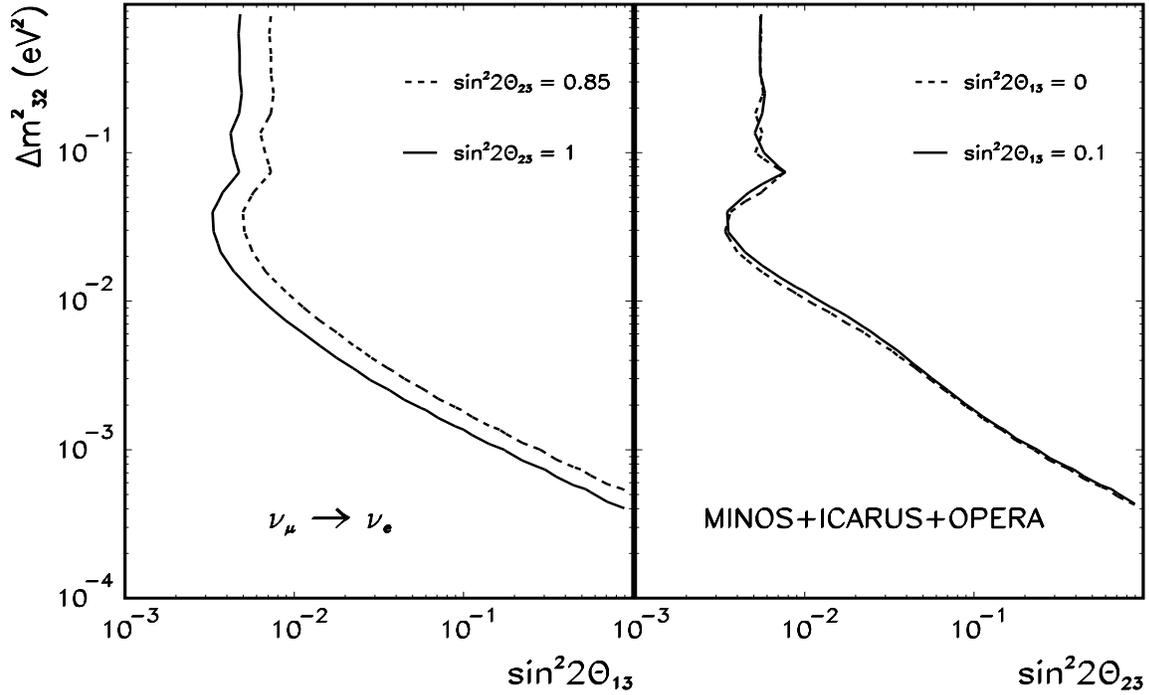,height=14.82cm,width=16.cm}}
\vglue -0.5cm
\caption{The global sensitivity 
of the $\nu_\mu \to \nu_e$ mode at
90\% C.L. is shown on the left. The sensitivity to $\sin^2 2\theta_{23}$ and 
$\Delta m^2_{32}$ by combining
the MINOS sensitivity in the disappearance channel and the ICARUS and OPERA sensitivities
in the $\nu_\mu \to \nu_\tau$ channels at the  90\% C.L. is on the right. 
}
\label{fig4a}
\vglue -1.cm
\end{figure}

\begin{figure}
\vglue -1.5cm
\hglue -0.5cm
\centerline{
\epsfig{file= 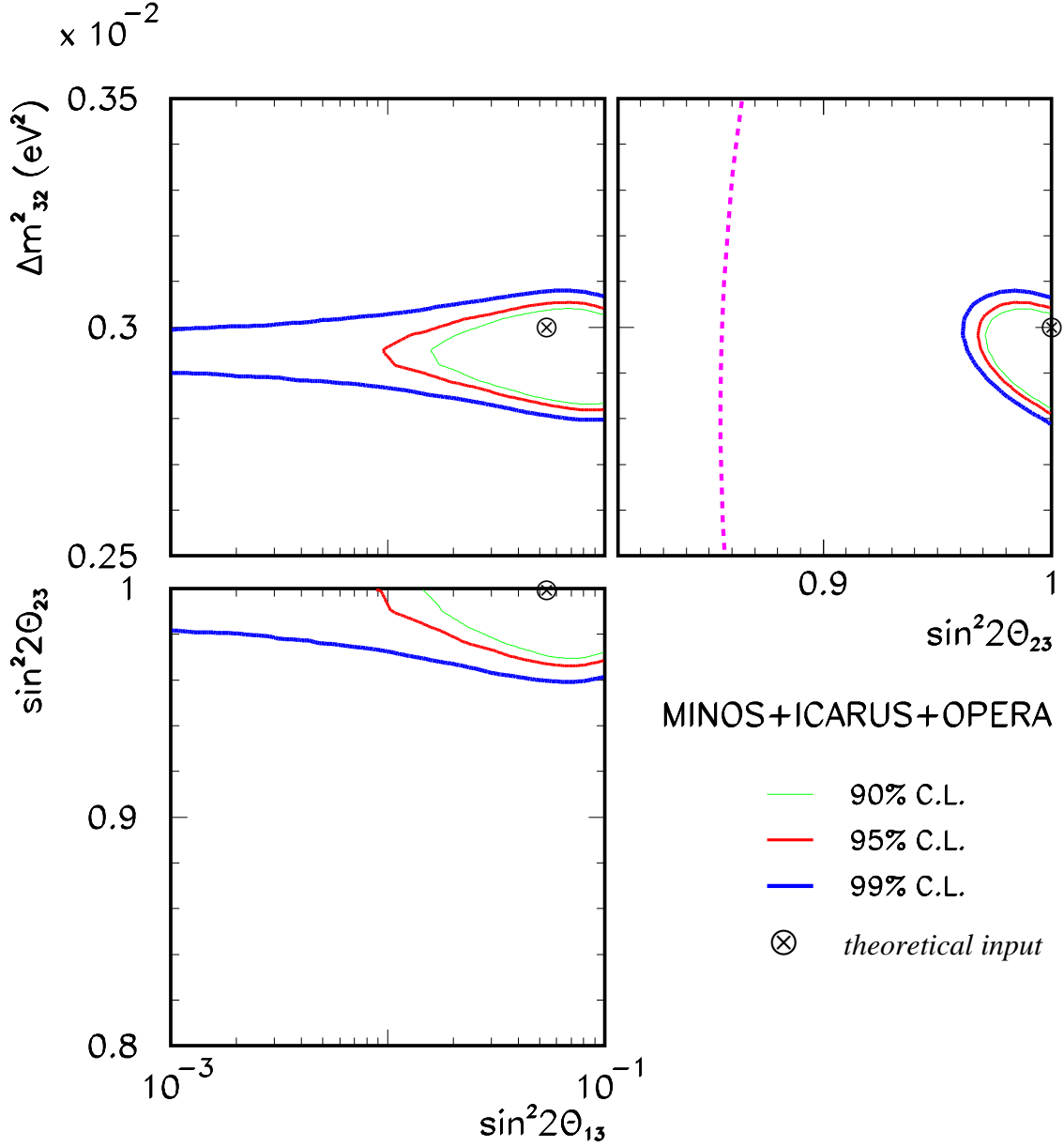 ,height=25.0cm,width=18.5cm}}
\vglue -4.5cm
\caption{
Expected allowed regions from the cumulative simulated data of the three experiments 
at the 90\%, 95\% and 99\% C.L. using
$\Delta m^2_{32}= 3 \times 10^{-3}$ eV$^2$,  
$\Delta m^2_{21}= 5 \times 10^{-5}$ eV$^2$, 
$\sin^2 2 \theta_{23} =  1$,   
$\sin^2 2 \theta_{12} = 0.8$ and
$\sin^2 2 \theta_{13} = 0.05$ as the theoretical input.
The dashed line is the Super-Kamiokande 
allowed region at the 99\% C.L..
}
\label{fig4b}
\vglue -1.cm
\end{figure}

The range of values of $\Delta m^2_{32}$ and $\sin^2 2\theta_{23}$ allowed 
by the Super-Kamiokande atmospheric neutrino data 
will be significantly narrowed
by the MINOS experiment. With an optimal $\langle L/E\rangle$ ratio for studying
these parameters, MINOS should pin them down to within 10\% at the 99\% C.L. with 10
kton-years of data. See Fig.~\ref{fig1b}.
ICARUS and OPERA, though not as sensitive to these parameters as MINOS,
will provide an important check on the 
neutrino oscillation hypothesis by observing tau events via 
the $\nu_\mu \to \nu_\tau$ channel. The value of $\sin^2 2\theta_{13}$ will be difficult for
these experiments to determine because of low
event rates and large backgrounds in the $\nu_e$ channel. We
have shown that 
this parameter will remain unbounded 
when the experiments are analysed separately. Combining the data from the MINOS and
ICARUS experiments may allow a lower bound on
$\sin^2 2\theta_{13}$ of $\cal{O}$(0.01) to be placed at the 95\% C.L. 
if $\sin^2 2\theta_{13}$ lies within the combined
sensitivity of the experiments. Establishing a lower bound on $\sin^2 2\theta_{13}$ would 
eliminate models
that predict smaller values of $\sin^2 2\theta_{13}$. 
If $\sin^2 2\theta_{13}>0.01$ and the $CP$ phase is sufficiently large, 
$CP$ violation in the lepton sector could be studied at superbeam facilities~\cite{bmw}.
  
In the next decade, with data from the three 
experiments considered (and a little luck), we could have good knowledge of the parameters
that mediate atmospheric neutrino oscillations. Along with KamLAND's
possible determination of $\Delta m^2_{12}$ and $\sin^2 2\theta_{12}$ using
reactor neutrinos~\cite{bdw}, all the elements of the mixing matrix other than the
$CP$ phase could be known.


\acknowledgments 
We thank A. Bettini for a critical reading of the manuscript and 
A. Erwin, M. Goodman and D. Harris for discussions. 
We are grateful to  M.\ Messier for providing us with the MINOS 
energy spectrum for the low energy beam
and to D.\ A.\ Petyt for information on MINOS efficiencies. 
\mbox{D. M.} and B. P. W. thank Universidade de S\~ao Paulo for its hospitality
during the initial stages of this work.
This work was supported by Funda\c{c}\~ao de Amparo \`a Pesquisa
do Estado de S\~ao Paulo (FAPESP), by Conselho Nacional de e 
Ci\^encia e Tecnologia (CNPq), by a U.S. NSF grant INT-9805573, by 
Department of Energy grants DE-FG02-91ER40676 and
DE-FG02-95ER40896 and by the University of Wisconsin Research Committee with
funds granted by the Wisconsin Alumni Research Foundation.


%

%
%

%
%

%
%

%
%

%
%

%
%

%
%

%
%

\end{document}